\documentclass[conference]{IEEEtran}
\IEEEoverridecommandlockouts
\usepackage{cite}
\usepackage{amsmath,amssymb,amsfonts,bm}

\usepackage{tabularx}
\usepackage{booktabs}
\usepackage{array}
\newcommand{\heading}[1]{\multicolumn{1}{c}{#1}}

\usepackage{graphicx}

\ifCLASSOPTIONcompsoc
    \usepackage[caption=false, font=normalsize, labelfont=sf, textfont=sf]{subfig}
\else
\usepackage[caption=false, font=footnotesize]{subfig}
\fi
\usepackage{xcolor}

\usepackage{tikz}
\usetikzlibrary{
    calc,
    chains,
    shapes, shapes.geometric,
    arrows, arrows.meta
}

\usepackage{pgfplots}
\pgfplotsset{compat=1.14}
\usepgfplotslibrary{fillbetween}
\usepackage{pgfplotstable}
\pgfplotstableset{col sep=comma}

\usepackage{siunitx}

\definecolor{set1-darkred}{RGB}{215,48,39}
\definecolor{set1-lightred}{RGB}{252,141,89}
\definecolor{set1-verylightred}{RGB}{254,224,144}
\definecolor{set1-verylightblue}{RGB}{224,243,248}
\definecolor{set1-lightblue}{RGB}{145,191,219}
\definecolor{set1-darkblue}{RGB}{69,117,180}

\begin{document}

\onecolumn

\noindent \textcopyright{} 2019 IEEE. Personal use of this material is permitted. Permission from IEEE must be obtained for all
other uses, in any current or future media, including reprinting/republishing this material for advertising or
promotional purposes, creating new collective works, for resale or redistribution to servers or lists, or reuse
of any copyrighted component of this work in other works.

\twocolumn

\title{%
Machine Learning to Tackle the Challenges of Transient and Soft Errors in Complex Circuits
\thanks{This work was supported by the RESCUE project which has received funding from the European Union's Horizon 2020 research and innovation programme under the Marie Sklodowska-Curie grant agreement No. 722325.}
}

\author{%
\IEEEauthorblockN{%
  Thomas Lange\IEEEauthorrefmark{1}\IEEEauthorrefmark{2},
  Aneesh Balakrishnan\IEEEauthorrefmark{1}\IEEEauthorrefmark{3},
  Maximilien Glorieux\IEEEauthorrefmark{1},
  Dan Alexandrescu\IEEEauthorrefmark{1},
  Luca Sterpone\IEEEauthorrefmark{2}%
}
\IEEEauthorblockA{%
  \IEEEauthorrefmark{1}\textit{iRoC Technologies}, Grenoble, France \\
  \IEEEauthorrefmark{2}\textit{Dipartimento di Informatica e Automatica, Politecnico di Torino}, Torino, Italy \\
  \IEEEauthorrefmark{3}\textit{Department of Computer Systems, Tallinn University of Technology}, Tallinn, Estonia \\
  \{thomas.lange, aneesh.balakrishnan, maximilien.glorieux, dan.alexandrescu\}@iroctech.com \qquad
  luca.sterpone@polito.it}
}

\maketitle

\begin{abstract}

The Functional Failure Rate analysis of today's complex circuits is a difficult task and requires a significant investment in terms of human efforts, processing resources and tool licenses. Thereby, de-rating or vulnerability factors are a major instrument of failure analysis efforts. Usually computationally intensive fault-injection simulation campaigns are required to obtain a fine-grained reliability metrics for the functional level. Therefore, the use of machine learning algorithms to assist this procedure and thus, optimising and enhancing fault injection efforts, is investigated in this paper. Specifically, machine learning models are used to predict accurate per-instance Functional De-Rating data for the full list of circuit instances, an objective that is difficult to reach using classical methods. The described methodology uses a set of per-instance features, extracted through an analysis approach, combining static elements (cell properties, circuit structure, synthesis attributes) and dynamic elements (signal activity). Reference data is obtained through first-principles fault simulation approaches. One part of this reference dataset is used to train the machine learning model and the remaining is used to validate and benchmark the accuracy of the trained tool. The presented methodology is applied on a practical example and various machine learning models are evaluated and compared.

\end{abstract}


\section{Introduction}

Due to technology scaling, lower supply voltages and higher operating frequencies, modern circuits become more and more vulnerable to reliability threats. Additionally, today's reliability standards and customers' expectations set tough targets for the quality of electronic devices and systems. Especially, transient faults, such as Single-Event Upsets and Single-Event Transients in the individual sequential and combinatorial cells, have been identified as the leading contributor to the overall failure rate for many applications~\cite{baumann_radiation-induced_2005, seifert_radiation-induced_2006}. Therefore, determining the Soft-Error Rate of the circuit is an important task. However, due to the increasing complexity of today's circuits, a detailed failure analysis requires a significant investment in terms of human efforts, processing resources and tool licenses. Therefore, new methodologies need to be considered, in order to lower the cost of failure analysis efforts.

\subsection{Objective of Our Methodology}

One of the major metrics used in today's functional safety analysis are de-rating or vulnerability factors. Using classical methods to obtain accurate per-instance Functional De-Rating data for the full list of circuit instances is a complex and computationally intensive task. Therefore, the methodology used in this paper assists this procedure with the help of machine learning algorithms. Previous works have shown that certain characteristics of the circuit, such as structure and signal probability, can be related to the masking effect and thus, used to estimate vulnerability factors~\cite{wali_low-cost_2017, ruano_methodology_2009, samudrala_selective_2004}. Since machine learning algorithms are very suitable to learn even complex relationships, we expect that these models are able to learn and predict the Functional De-Rating by using similar circuit features. The here used set of features, which individually characterises each flip-flop instance in the circuit, is used to train the machine learning model in a supervised learning approach. The trained model predicts the remaining Functional De-Rating values for the flip-flop instances, which were not used for training. The methodology is applied on a practical example and the paper focuses on the evaluation of various machine learning models in terms of performance and timing.

\subsection{Organisation of the Paper}

The rest of this paper is organised as follows: Section~\ref{sec:background} briefly summarises the definition of Single-Event Effects and the different de-rating mechanism. The methodology is described in section~\ref{sec:methodology} and applied on a practical example in section~\ref{sec:results}. Various machine learning models are evaluated and compared to each other. Section~\ref{sec:conclusion} summarises this paper and gives concluding remarks, as well as prospects for future work.

\section{De-Rating Mechanisms}
\label{sec:background}

Transient faults, or Single-Event Effects (SEE), are caused by a single, energetic particles striking through the device. The subsequent propagation of this fault in the system can lead to observable effects (failures) up to the system level. However, not all faults necessarily manifest themselves as errors or failures in the system. Four de-rating mechanisms can reduce the effect of SEEs on the actual error rate significantly~\cite{nguyen_systematic_2003, alexandrescu_towards_2012}. \emph{Electrical De-Rating (EDR)} describes the effect when the transient pulse is filtered due to narrowing. By the time the transient reaches the end of the path the voltage of the pulse is below the switching threshold. \emph{Temporal De-Rating (TDR)} applies when the transient fault reaches the input of a sequential element but is not sampled, because it arrives outside the latching window. \emph{Logical De-Rating (LDR)} means that the transient fault is not propagating due to the state of a controlling input of a gate (e.g. value \verb+1+ on an \verb+OR2+ gate). The \emph{Functional De-Rating (FDR)} considers the transient fault on the application level. Even when the transient fault is not masked by any other mechanism and thus, propagates to the system output, the impact at the functional level can considerably vary. Depending on the criteria defining the acceptable behaviour of the circuit, in many cases the fault is benign. 

These de-rating mechanisms are used to evaluate the probability of the propagation of a fault and are usually determined by using probabilistic algorithms and simulation based approaches. All the different evaluation steps can require significant investment in terms of human efforts, processing resources and licenses for different tools. Thereby, especially the simulation based approaches to determine the Functional De-Rating are very computationally intensive.

The used methodology in this paper estimates the Functional De-Rating factors for individual flip-flops with the help of Machine Learning algorithms. Therefore, supervised regression models are used to predict the continues variable. These models try to learn the dependency between a set of input features and the target output variable, usually based upon a mathematical model, by using sample data (also called training data). The learned relationships is used to predict new data points~\cite{alpaydin_introduction_2014}. The methodology is presented in detail in the next section.

\section{Methodology}
\label{sec:methodology}

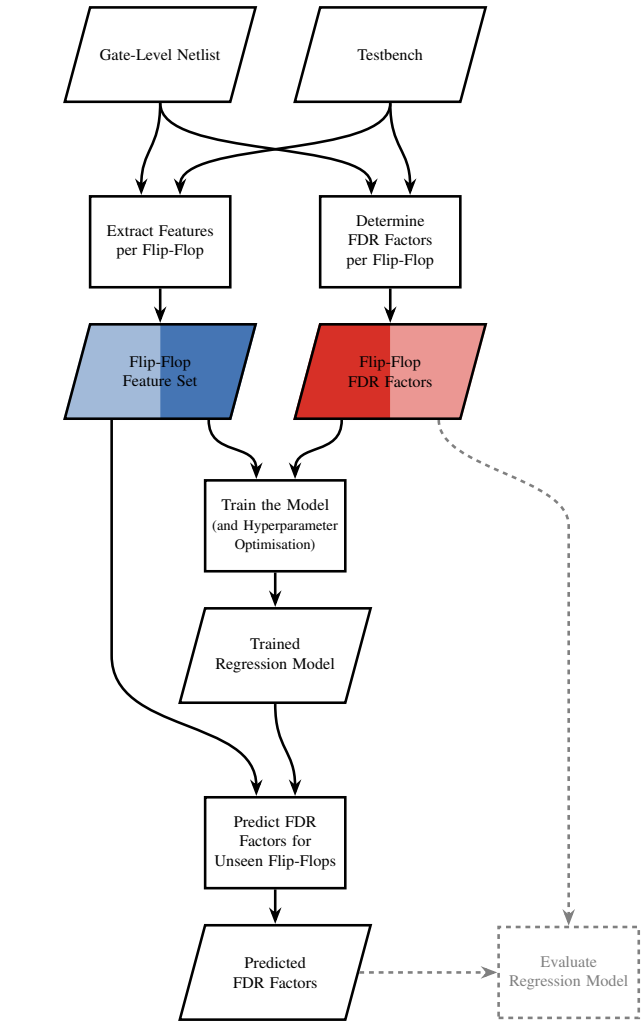
\begin{figure}[htbp]
    \centering
    \resizebox{0.9\linewidth}{!}{\newcommand{\boxit}[1]{\parbox[c][11.5ex][c]{7.5em}{\centering #1}}

\pgfdeclarelayer{background}
\pgfsetlayers{background,main}

\begin{tikzpicture}[%
    >={Stealth[scale=1]},              %
    start chain=going below,    %
    node distance=7.5mm and 50mm, %
    every join/.style={norm},   %
    ]
\tikzset{
  base/.style={draw, ultra thick, 
    on chain, on grid, align=center, minimum height=12.5ex
  },
  proc/.style={base, rectangle, text width=8em},
  term/.style={proc, rounded corners},
  test/.style={base, diamond, aspect=2, text width=5em},
  data/.style={base, trapezium, trapezium stretches=true,
    trapezium left angle=75, trapezium right angle=-75,
    text width=8em,
    node contents={\boxit{#1}}
  },
  coord/.style={coordinate, on chain, on grid, node distance=6mm and 25mm},
  nmark/.style={draw, cyan, circle, font={\sffamily\bfseries}},
  norm/.style={->, ultra thick, draw},
  it/.style={font={\small\itshape}}
}

\node (gln) [data=Gate-Level Netlist];
\node [proc, below=20mm of gln.south] (ext-f) {Extract Features per Flip-Flop};
\node (ff-fset) [data=Flip-Flop \\ Feature Set];

\node (tb) [data=Testbench, right=of gln];
\node [proc, below=20mm of tb.south] (ext-fdr) {Determine FDR Factors per Flip-Flop};
\node (ff-fdr) [data=Flip-Flop \\ FDR Factors];

\node [coord] (flowhmid) at ($(ext-f)!0.5!(ff-fdr)$) {};
\node [coord] {};

\node [proc, below=30mm of flowhmid] (train) {Train the Model \small (and Hyperparameter Optimisation)};
\node (trained-model) [data=Trained Regression Model];
\node [proc, below=20mm of trained-model.south] (estimate) {Predict FDR Factors for Unseen Flip-Flops};

\node (estimated-fdr) [data=Predicted \\ FDR Factors];

\node [proc, dashed, gray, right=30mm of estimated-fdr.east] (evaluate) {Evaluate Regression Model};

\draw [norm] (gln) edge[out=-90, in=90, looseness=1.25] (ext-f.112.5);
\draw [norm] (gln) edge[out=-90, in=90, looseness=0.625] (ext-fdr.112.5);

\draw [norm] (tb) edge[out=-90, in=90, looseness=0.625] (ext-f.67.5);
\draw [norm] (tb) edge[out=-90, in=90, looseness=1.25] (ext-fdr.67.5);

\draw [norm] (ext-f) -- (ff-fset);
\draw [norm] (ext-fdr) -- (ff-fdr);

\draw [norm] (ff-fset.315) to[out=-90, in=90, looseness=1.25] (train.112.5);
\draw [norm] (ff-fdr.225) to[out=-90, in=90, looseness=1.25] (train.67.5);

\draw [norm] (train) -- (trained-model);

\draw [norm] (trained-model) edge[out=-90, in=90, looseness=1.25] (estimate.67.5);
\draw [norm] (ff-fset.225) 
    to[out=-90, in=90, looseness=1] (ff-fset.225 |- train)
    -- (ff-fset.225 |- trained-model)
    to[out=-90, in=90, looseness=1] (estimate.112.5);

\draw [norm]  (estimate) -- (estimated-fdr);

\draw [norm, dashed, gray]  (estimated-fdr) -- (evaluate);
\draw [norm, dashed, gray]  (ff-fdr.315)
    to[out=-90, in=90, looseness=0.75] (train -| evaluate) 
    -- (evaluate);

\begin{pgfonlayer}{background}
\filldraw [draw=none, fill=set1-darkblue, opacity=0.5] 
    (ff-fset.top left corner) -- (ff-fset.north) --
    (ff-fset.south) -- (ff-fset.bottom left corner) -- cycle;
    
\filldraw [draw=none, fill=set1-darkblue] 
    (ff-fset.top right corner) -- (ff-fset.north) --
    (ff-fset.south) -- (ff-fset.bottom right corner) -- cycle;
    
\filldraw [draw=none, fill=set1-darkred, opacity=1] 
    (ff-fdr.top left corner) -- (ff-fdr.north) --
    (ff-fdr.south) -- (ff-fdr.bottom left corner) -- cycle;
    
\filldraw [draw=none, fill=set1-darkred, opacity=0.5] 
    (ff-fdr.top right corner) -- (ff-fdr.north) --
    (ff-fdr.south) -- (ff-fdr.bottom right corner) -- cycle;
\end{pgfonlayer}

\end{tikzpicture}}
    \caption{Functional De-Rating Estimation and Evaluation Flow}
    \label{fig:ML_flow}
\end{figure}

The implemented procedure to predict the Functional De-Rating factors per flip-flop instance is based on machine learning regression models and shown in Fig.~\ref{fig:ML_flow}. The gate-level netlist of the circuit and a corresponding testbench are used to extract the features for each flip-flop in the circuit. They are also used in a statistical fault injection simulation to determine the FDR factors for one part of the circuit. The determined FDR factors per flip-flop and the associated flip-flop features form the training data set, used to train the ML model. The size of the training data set is defined by the training size and thus, also defines the number of fault injections to perform. Eventually, the trained model can be used to estimate the FDR values of the remaining flip-flops.

\subsection{Flip-Flop Feature Extraction}

The extraction of the flip-flop feature set needs to be efficient in order to compete with the classical fault injection approach. Therefore, the used feature set to characterise each flip-flop instance, contains only simple characteristics which are easy and fast to obtain. The feature set combines static elements, such as cell properties, circuit structure and synthesis attributes, as well as dynamic elements, such as signal activity.

The structural features for each flip-flop are flip-flop fan-in/fan-out, number of connections from/to other flip-flops within the circuit, connections from/to primary input/output and the corresponding proximity (number of stages), if the flip-flop is part of a bus, the bus position and corresponding bus length, and if the flip-flops has a feedback loop and the corresponding depth (number of stages). Further features were extracted which are related to the synthesis of the circuit. The synthesis related features are flip-flop drive strength, combinatorial fan-in/fan-out and the depth of the combinatorial path. To consider the workload of the circuit, features are required which describe the dynamic behaviour of the flip-flops. Therefore, the signal activity for each flip-flop is described by the number of state changes and the time the output is at logical~\verb+0+/\verb+1+.

\subsection{Model Training and Hyperparameter Optimisation}

Machine learning models are usually represented by internal parameters or an internal state. These parameters or the state are determined during the training process by the machine learning algorithm. Additionally, most of the machine learning algorithms can be controlled by hyperparameters. In contrast to the internal parameters or state, these hyperparameters are not derived by the training algorithm and need to be manually set before the training process. The problem of finding the optimal set hyperparameters for the model is called hyperparameter optimisation. Therefore, several instances of the model need to be trained and evaluated for different tuples of hyperparameters. The tuple that minimises a predefined loss function or evaluation metrics yields an optimal model. A random search method combined with a grid search method is a common approach to perform the optimisation. There, the model is first evaluated for parameter values randomly generated in a given distribution. Afterwards a more detailed grid search is performed within the region of the values obtained by the random search~\cite{bergstra_random_2012}.

\subsection{Model Evaluation}

In order to measure and evaluate the performance of a machine learning model different metrics are used. In this paper the mean absolute error (MAE), the maximum absolute error (MAX), the root-mean-square error (RMSE), the explained variance (EV) and the coefficient of determination ($R^2$) are considered. Further, cross-validation is used to ensure that the model is not only trained for one particular training and test data set. There, the model is trained and evaluated against multiple train and test splits of the data. Several subsets, or cross validation folds, of the data set are created and each fold is used to train and evaluate a separate model. Thus, instead of relying only on one single training and test data set, a more stable performance measure is obtained which indicates how the model is likely to perform on average~\cite{kohavi_study_1995}.

\colorlet{learn-train-color}{set1-lightred}
\colorlet{learn-test-color}{set1-lightblue}
\colorlet{learn-time-color}{gray}

\colorlet{example-true-color}{set1-darkblue}
\colorlet{example-predicted-color}{set1-darkred}

\pgfplotsset{
    scale only axis,
    base-plot/.style={
        cycle list name=mark list,
        xlabel shift=-0.325em,
        font=\scriptsize,
        tick label style={font=\tiny},
        legend cell align=left,
        legend style={
            font=\tiny, 
            fill opacity=0.75, 
            text opacity=1.0
        },
    },
    plot-learn/.style={
        base-plot,
        width=4.375cm, height=3.4cm,
        axis y line*=left,
        grid,
        xtick distance=20,
        xmin=0,
        xmax=100,
        xticklabel={\pgfmathprintnumber\tick\%},
        xlabel={Training Size},
        ymin=0,
        ymax=1.05,
        ytick distance=0.2,
        ylabel={$R^2$ Score},
        ylabel shift=-0.45em,
        legend columns=-1,
        legend style={at={(0.5,1.02)},anchor=south}
    },
    plot-learn-time/.style={
        base-plot,
        width=4.375cm, height=3.4cm,
        axis x line=none,
        axis y line*=right,
        xtick distance=20,
        xmin=0,
        xmax=100,
        xticklabel={\pgfmathprintnumber\tick\%},
        ymin=0,
        ymax=105,
        ytick distance=20,
        ylabel={Fit Time / ms},
        ylabel shift=-0.75em,
    },
    plot-example/.style={
        base-plot,
        width=5.075cm, height=4.2cm,
        grid,
        only marks,
        mark size=0.75,
        xtick distance=100,
        enlarge x limits={abs=10},
        xlabel={Flip-Flop $FF_i$},
        ymax=1.05,
        ymin=-0.05,
        ytick distance=0.2,
        ylabel={Functional De-Rating},
        ylabel shift=-0.45em,
        legend pos=south east,
    },
    plot-scatter/.style={
        base-plot,
        xmin=-0.05, xmax=1.05,
        xtick distance=0.2,
        xlabel={FDR Test True Values},
        ymin=-0.05, ymax=1.05,
        ytick distance=0.2,
        ylabel={FDR Test Predicted Values},
        ylabel shift=-0.125em
    },
}
\section{Evaluating Machine Learning Models Predicting Functional De-Rating Factors}
\label{sec:results}

In this section the presented methodology is applied on a practical example and various machine learning models are evaluated and compared. Therefore, first, a full flat statistical fault injection campaign was performed to get the Functional De-Rating factors for each flip-flop. These are partly used to train the machine learning models and further, serve as a reference for the model evaluation. Afterwards, the trained models are used to predict the individual Functional De-Rating factors. The performance of the prediction is evaluated for different training sizes by using several metrics. Additionally, the fit and prediction time for different training sizes were measured and compared to each other, as well as the time to perform the hyperparameter optimisation, which is compared against the full fault injection campaign\footnote{All computations were performed on a PC with an Intel Xeon E5-2687W CPU (\SI{8}{cores}/\SI{16}{threads} \@ \SI{3.10}{GHz}).}.

\subsection{Circuit Under Test}

For the practical example, the Ethernet 10GE~MAC Core from OpenCores is used. This circuit implements the Media Access Control (MAC) functions as defined in the IEEE~802.3ae standard. The 10GE~MAC core has a 10\,Gbps interface (XGMII TX/RX) to connect it to different types of Ethernet PHYs and one packet interface to transmit and receive packets to/from the user logic~\cite{andre_tanguay_10ge_2013}. The circuit consists of control logic, state machines, FIFOs and memory interfaces. It is implemented at the Register-Transfer Level (RTL) and is publicly available on OpenCores.

The corresponding testbench writes several packets to the 10GE~MAC transmit packet interface. As packet frames become available in the transmit FIFO, the MAC calculates a CRC and sends them out to the XGMII transmitter. The XGMII~TX interface is looped-back to the XGMII~RX interface in the testbench. The frames are thus processed by the MAC receive engine and stored in the receive FIFO. Eventually, the testbench reads frames from the packet receive interface and prints out the results~\cite{andre_tanguay_10ge_2013}. During the simulation all sent and received packages to and from the core are monitored and recorded. This record is used as the golden reference for the fault injection campaign.

By synthesising the design using the NanGate FreePDK45 Open Cell Library~\cite{stine_freepdk_2007}, the gate-level netlist was obtained and 1054\,flip-flops have been identified. The netlist and the testbench are used to extract the respective features for each flip-flop, which took \SI{184}{seconds}.

\subsection{Failure Classes and Fault Injection Campaign}
\label{sec:fi_campaign}

In order to obtain the sensitivity of each flip-flop and thus, determine the Functional De-Rating, a flat statistical fault injection campaign was performed on the gate-level netlist. One the one hand, part of this dataset is used to train the machine learning models. On the other hand, it provides an objective measure and the part of the dataset which was not used for training, is used to evaluate the trained models.

For each of the 1054 flip-flops 170 faults were injected at a random time during the active phase of the simulation. The fault injection mechanism is implemented by inverting the value stored in a flip-flop using a simulator function. In total \SI{16}{hours} \SI{41}{minutes} and \SI{20}{seconds} were needed to perform the campaign.

For the analysis two different failure classes are considered. In case the injected fault propagates to the primary outputs of the circuit and thus, the output values are altered in comparison to the golden reference, an \emph{Output Failure} is counted. Further, if the payload of the final received packages is corrupted or the circuit stopped sending or receiving data, the simulation run was considered as an \emph{Application Failure}. Eventually, the corresponding De-Rating factor is calculated by the number of simulation runs with an Output Failure or Application Failure respectively, divided by the number of total simulation runs.

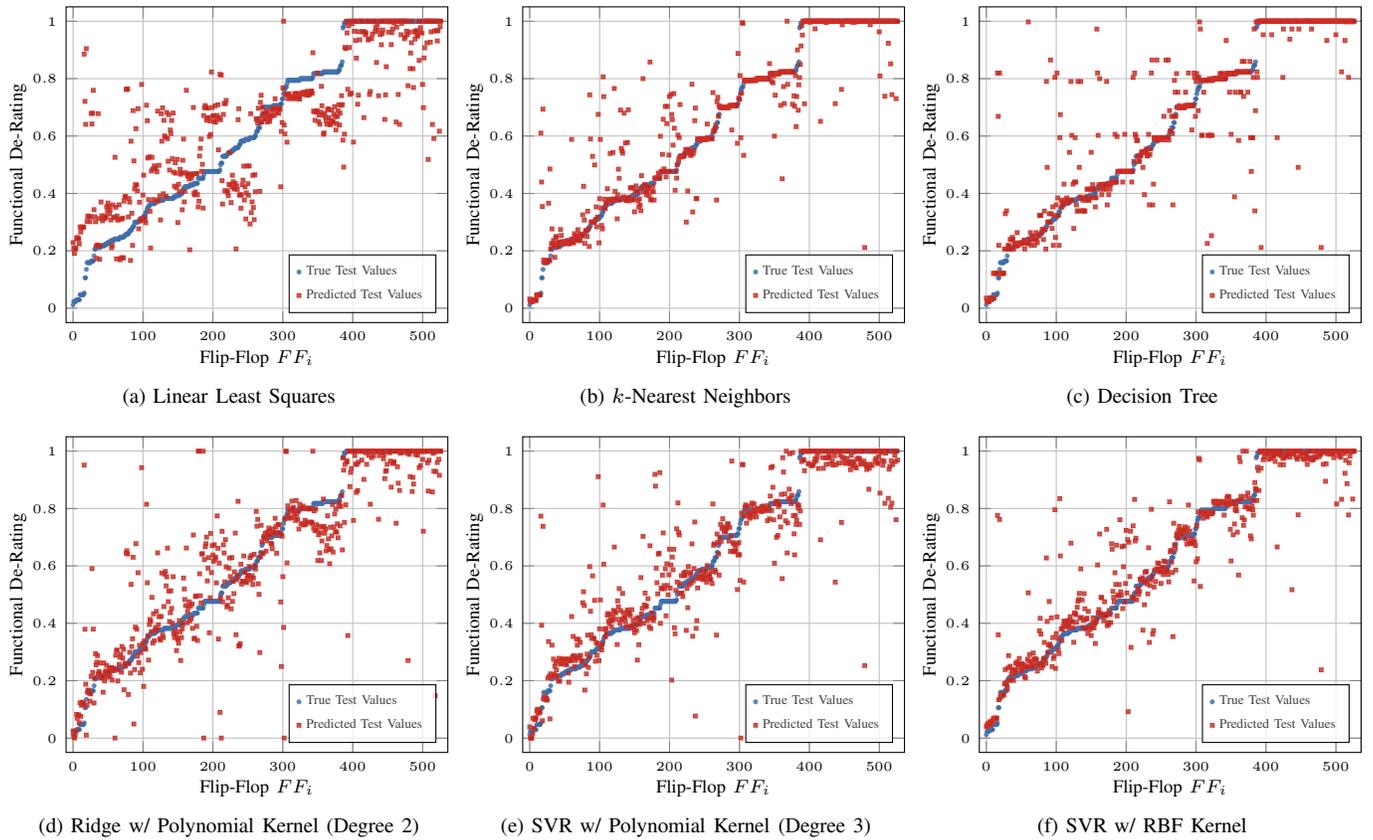
\begin{figure*}[htbp]
    \centering
    \subfloat[Linear Least Squares]{%
        \pgfplotstableread{fig/out_err/LinearRegression_example.csv}\examplecsv%
        \begin{tikzpicture}
    \begin{axis}[
        plot-example
    ]
        \addplot+ [example-true-color, opacity=0.75] 
            table [x=FF idx, y=test true]{\examplecsv};
        \addplot+ [color=example-predicted-color, opacity=0.75]
            table [x=FF idx, y=test predicted]{\examplecsv};
        \legend{True Test Values, Predicted Test Values}
    \end{axis}
\end{tikzpicture}%
    }%
    \subfloat[$k$-Nearest Neighbors]{%
        \pgfplotstableread{fig/out_err/KNeighborsRegressor_example.csv}\examplecsv%
        \begin{tikzpicture}
    \begin{axis}[
        plot-example
    ]
        \addplot+ [example-true-color, opacity=0.75] 
            table [x=FF idx, y=test true]{\examplecsv};
        \addplot+ [color=example-predicted-color, opacity=0.75]
            table [x=FF idx, y=test predicted]{\examplecsv};
        \legend{True Test Values, Predicted Test Values}
    \end{axis}
\end{tikzpicture}%
    }%
    \subfloat[Decision Tree]{%
        \pgfplotstableread{fig/out_err/DecisionTreeRegressor_example.csv}\examplecsv%
        \begin{tikzpicture}
    \begin{axis}[
        plot-example
    ]
        \addplot+ [example-true-color, opacity=0.75] 
            table [x=FF idx, y=test true]{\examplecsv};
        \addplot+ [color=example-predicted-color, opacity=0.75]
            table [x=FF idx, y=test predicted]{\examplecsv};
        \legend{True Test Values, Predicted Test Values}
    \end{axis}
\end{tikzpicture}%
    }%
    
    \subfloat[Ridge w/ Polynomial Kernel (Degree 2)]{%
        \pgfplotstableread{fig/out_err/KernelRidgeRegressor_2DegreePolyKernel_example.csv}\examplecsv%
        \begin{tikzpicture}
    \begin{axis}[
        plot-example
    ]
        \addplot+ [example-true-color, opacity=0.75] 
            table [x=FF idx, y=test true]{\examplecsv};
        \addplot+ [color=example-predicted-color, opacity=0.75]
            table [x=FF idx, y=test predicted]{\examplecsv};
        \legend{True Test Values, Predicted Test Values}
    \end{axis}
\end{tikzpicture}%
    }%
    \subfloat[SVR w/ Polynomial Kernel (Degree 3)]{%
        \pgfplotstableread{fig/out_err/SVR_PolyKernel_example.csv}\examplecsv%
        \begin{tikzpicture}
    \begin{axis}[
        plot-example
    ]
        \addplot+ [example-true-color, opacity=0.75] 
            table [x=FF idx, y=test true]{\examplecsv};
        \addplot+ [color=example-predicted-color, opacity=0.75]
            table [x=FF idx, y=test predicted]{\examplecsv};
        \legend{True Test Values, Predicted Test Values}
    \end{axis}
\end{tikzpicture}%
    }%
    \subfloat[SVR w/ RBF Kernel]{%
        \pgfplotstableread{fig/out_err/SVR_RBFKernel_example.csv}\examplecsv%
        \begin{tikzpicture}
    \begin{axis}[
        plot-example
    ]
        \addplot+ [example-true-color, opacity=0.75] 
            table [x=FF idx, y=test true]{\examplecsv};
        \addplot+ [color=example-predicted-color, opacity=0.75]
            table [x=FF idx, y=test predicted]{\examplecsv};
        \legend{True Test Values, Predicted Test Values}
    \end{axis}
\end{tikzpicture}%
    }%
    \caption{Prediction of the Output Failure for one test data fold (training size = 50\%).}
    \label{fig:out-err-example}
\end{figure*}

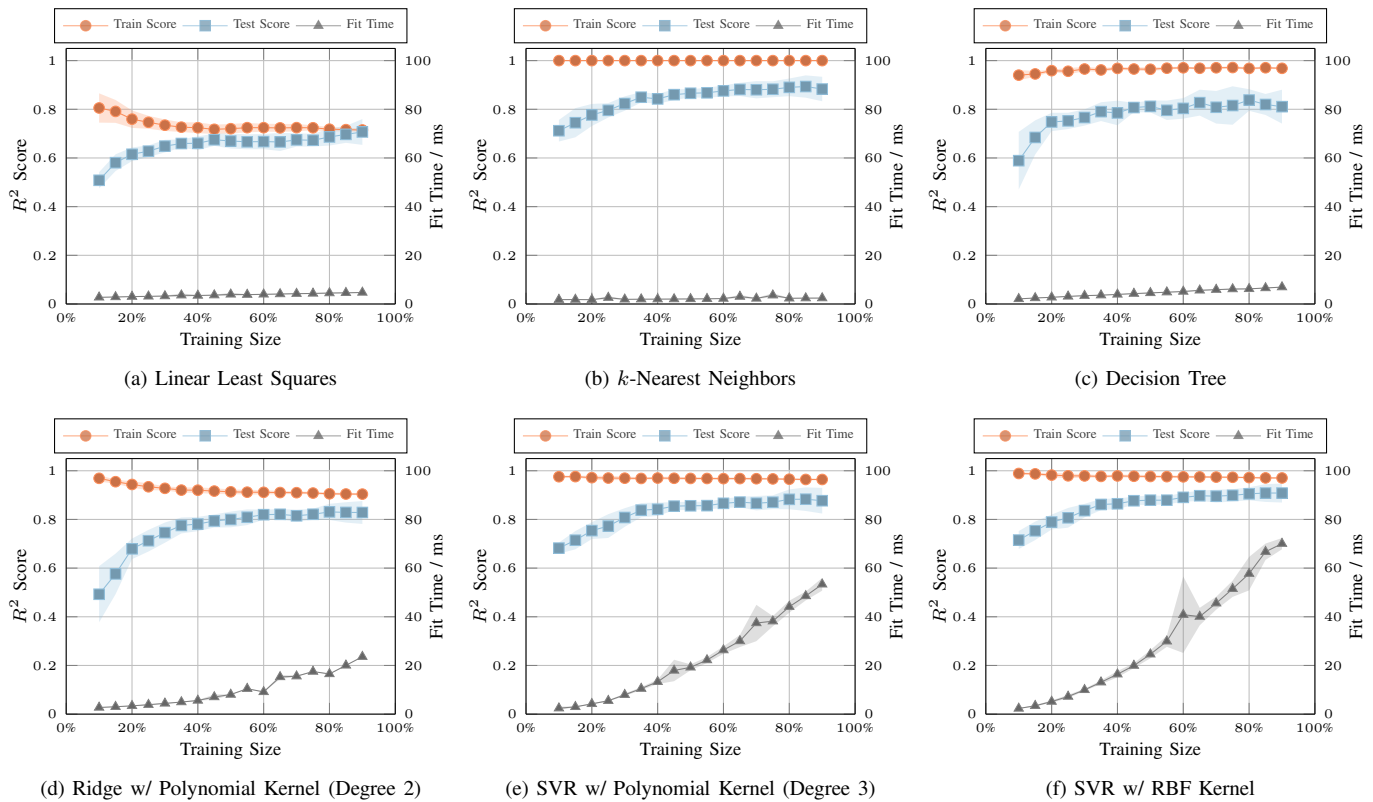
\begin{figure*}[htbp]
    \vspace{-20pt}%
    \centering%
    \subfloat[Linear Least Squares]{%
        \pgfplotstableread{fig/out_err/LinearRegression_learn.csv}\learncsv%
        \begin{tikzpicture}
    \begin{axis}[
        plot-learn-time
    ]
        \addplot+ [learn-time-color, mark=triangle*]
            table [
                x expr=\thisrow{train size}*100,
                y expr=\thisrow{fit time mean}*1000
            ]{\learncsv}; \label{fit-time}
        
        \addplot [name path=time upper, draw=none] 
            table [
                x expr=\thisrow{train size}*100,
                y expr=\thisrow{fit time mean}*1000+\thisrow{fit time std}*1000
            ] {\learncsv};
        \addplot [name path=time lower, draw=none] 
            table[
                x expr=\thisrow{train size}*100,
                y expr=\thisrow{fit time mean}*1000-\thisrow{fit time std}*1000
            ] {\learncsv};
        \addplot [fill=learn-time-color, opacity=0.25] fill between[of=time upper and time lower];
        
    \end{axis}
    
    \begin{axis}[
        plot-learn,
    ]
        \addplot+ [learn-train-color]
            table [
                x expr=\thisrow{train size}*100,
                y=train mean
            ]{\learncsv};
        \addplot+ [learn-test-color]
            table [
                x expr=\thisrow{train size}*100, 
                y=test mean
            ]{\learncsv};
        \legend{Train Score, Test Score}
        \addlegendimage{/pgfplots/refstyle=fit-time}\addlegendentry{Fit Time}
        
        \addplot [name path=train upper, draw=none] 
            table [
                x expr=\thisrow{train size}*100,
                y expr=\thisrow{train mean}+\thisrow{train std}
            ] {\learncsv};
        \addplot [name path=train lower, draw=none] 
            table[
                x expr=\thisrow{train size}*100,
                y expr=\thisrow{train mean}-\thisrow{train std}
            ] {\learncsv};
        \addplot [fill=learn-train-color, opacity=0.25] fill between[of=train upper and train lower];
        
        \addplot [name path=train upper, draw=none] 
            table [
                x expr=\thisrow{train size}*100,
                y expr=\thisrow{test mean}+\thisrow{test std}
            ] {\learncsv};
        \addplot [name path=train lower, draw=none] 
            table[
                x expr=\thisrow{train size}*100,
                y expr=\thisrow{test mean}-\thisrow{test std}
            ] {\learncsv};
        \addplot [fill=learn-test-color, opacity=0.25] fill between[of=train upper and train lower];
    \end{axis}
\end{tikzpicture}%
    }%
    \subfloat[$k$-Nearest Neighbors]{%
        \pgfplotstableread{fig/out_err/KNeighborsRegressor_learn.csv}\learncsv%
        \begin{tikzpicture}
    \begin{axis}[
        plot-learn-time
    ]
        \addplot+ [learn-time-color, mark=triangle*]
            table [
                x expr=\thisrow{train size}*100,
                y expr=\thisrow{fit time mean}*1000
            ]{\learncsv}; \label{fit-time}
        
        \addplot [name path=time upper, draw=none] 
            table [
                x expr=\thisrow{train size}*100,
                y expr=\thisrow{fit time mean}*1000+\thisrow{fit time std}*1000
            ] {\learncsv};
        \addplot [name path=time lower, draw=none] 
            table[
                x expr=\thisrow{train size}*100,
                y expr=\thisrow{fit time mean}*1000-\thisrow{fit time std}*1000
            ] {\learncsv};
        \addplot [fill=learn-time-color, opacity=0.25] fill between[of=time upper and time lower];
        
    \end{axis}
    
    \begin{axis}[
        plot-learn,
    ]
        \addplot+ [learn-train-color]
            table [
                x expr=\thisrow{train size}*100,
                y=train mean
            ]{\learncsv};
        \addplot+ [learn-test-color]
            table [
                x expr=\thisrow{train size}*100, 
                y=test mean
            ]{\learncsv};
        \legend{Train Score, Test Score}
        \addlegendimage{/pgfplots/refstyle=fit-time}\addlegendentry{Fit Time}
        
        \addplot [name path=train upper, draw=none] 
            table [
                x expr=\thisrow{train size}*100,
                y expr=\thisrow{train mean}+\thisrow{train std}
            ] {\learncsv};
        \addplot [name path=train lower, draw=none] 
            table[
                x expr=\thisrow{train size}*100,
                y expr=\thisrow{train mean}-\thisrow{train std}
            ] {\learncsv};
        \addplot [fill=learn-train-color, opacity=0.25] fill between[of=train upper and train lower];
        
        \addplot [name path=train upper, draw=none] 
            table [
                x expr=\thisrow{train size}*100,
                y expr=\thisrow{test mean}+\thisrow{test std}
            ] {\learncsv};
        \addplot [name path=train lower, draw=none] 
            table[
                x expr=\thisrow{train size}*100,
                y expr=\thisrow{test mean}-\thisrow{test std}
            ] {\learncsv};
        \addplot [fill=learn-test-color, opacity=0.25] fill between[of=train upper and train lower];
    \end{axis}
\end{tikzpicture}%
    }%
    \subfloat[Decision Tree]{%
        \pgfplotstableread{fig/out_err/DecisionTreeRegressor_learn.csv}\learncsv%
        \begin{tikzpicture}
    \begin{axis}[
        plot-learn-time
    ]
        \addplot+ [learn-time-color, mark=triangle*]
            table [
                x expr=\thisrow{train size}*100,
                y expr=\thisrow{fit time mean}*1000
            ]{\learncsv}; \label{fit-time}
        
        \addplot [name path=time upper, draw=none] 
            table [
                x expr=\thisrow{train size}*100,
                y expr=\thisrow{fit time mean}*1000+\thisrow{fit time std}*1000
            ] {\learncsv};
        \addplot [name path=time lower, draw=none] 
            table[
                x expr=\thisrow{train size}*100,
                y expr=\thisrow{fit time mean}*1000-\thisrow{fit time std}*1000
            ] {\learncsv};
        \addplot [fill=learn-time-color, opacity=0.25] fill between[of=time upper and time lower];
        
    \end{axis}
    
    \begin{axis}[
        plot-learn,
    ]
        \addplot+ [learn-train-color]
            table [
                x expr=\thisrow{train size}*100,
                y=train mean
            ]{\learncsv};
        \addplot+ [learn-test-color]
            table [
                x expr=\thisrow{train size}*100, 
                y=test mean
            ]{\learncsv};
        \legend{Train Score, Test Score}
        \addlegendimage{/pgfplots/refstyle=fit-time}\addlegendentry{Fit Time}
        
        \addplot [name path=train upper, draw=none] 
            table [
                x expr=\thisrow{train size}*100,
                y expr=\thisrow{train mean}+\thisrow{train std}
            ] {\learncsv};
        \addplot [name path=train lower, draw=none] 
            table[
                x expr=\thisrow{train size}*100,
                y expr=\thisrow{train mean}-\thisrow{train std}
            ] {\learncsv};
        \addplot [fill=learn-train-color, opacity=0.25] fill between[of=train upper and train lower];
        
        \addplot [name path=train upper, draw=none] 
            table [
                x expr=\thisrow{train size}*100,
                y expr=\thisrow{test mean}+\thisrow{test std}
            ] {\learncsv};
        \addplot [name path=train lower, draw=none] 
            table[
                x expr=\thisrow{train size}*100,
                y expr=\thisrow{test mean}-\thisrow{test std}
            ] {\learncsv};
        \addplot [fill=learn-test-color, opacity=0.25] fill between[of=train upper and train lower];
    \end{axis}
\end{tikzpicture}%
    }%
    
    \subfloat[Ridge w/ Polynomial Kernel (Degree 2)]{%
        \pgfplotstableread{fig/out_err/KernelRidgeRegressor_2DegreePolyKernel_learn.csv}\learncsv%
        \begin{tikzpicture}
    \begin{axis}[
        plot-learn-time
    ]
        \addplot+ [learn-time-color, mark=triangle*]
            table [
                x expr=\thisrow{train size}*100,
                y expr=\thisrow{fit time mean}*1000
            ]{\learncsv}; \label{fit-time}
        
        \addplot [name path=time upper, draw=none] 
            table [
                x expr=\thisrow{train size}*100,
                y expr=\thisrow{fit time mean}*1000+\thisrow{fit time std}*1000
            ] {\learncsv};
        \addplot [name path=time lower, draw=none] 
            table[
                x expr=\thisrow{train size}*100,
                y expr=\thisrow{fit time mean}*1000-\thisrow{fit time std}*1000
            ] {\learncsv};
        \addplot [fill=learn-time-color, opacity=0.25] fill between[of=time upper and time lower];
        
    \end{axis}
    
    \begin{axis}[
        plot-learn,
    ]
        \addplot+ [learn-train-color]
            table [
                x expr=\thisrow{train size}*100,
                y=train mean
            ]{\learncsv};
        \addplot+ [learn-test-color]
            table [
                x expr=\thisrow{train size}*100, 
                y=test mean
            ]{\learncsv};
        \legend{Train Score, Test Score}
        \addlegendimage{/pgfplots/refstyle=fit-time}\addlegendentry{Fit Time}
        
        \addplot [name path=train upper, draw=none] 
            table [
                x expr=\thisrow{train size}*100,
                y expr=\thisrow{train mean}+\thisrow{train std}
            ] {\learncsv};
        \addplot [name path=train lower, draw=none] 
            table[
                x expr=\thisrow{train size}*100,
                y expr=\thisrow{train mean}-\thisrow{train std}
            ] {\learncsv};
        \addplot [fill=learn-train-color, opacity=0.25] fill between[of=train upper and train lower];
        
        \addplot [name path=train upper, draw=none] 
            table [
                x expr=\thisrow{train size}*100,
                y expr=\thisrow{test mean}+\thisrow{test std}
            ] {\learncsv};
        \addplot [name path=train lower, draw=none] 
            table[
                x expr=\thisrow{train size}*100,
                y expr=\thisrow{test mean}-\thisrow{test std}
            ] {\learncsv};
        \addplot [fill=learn-test-color, opacity=0.25] fill between[of=train upper and train lower];
    \end{axis}
\end{tikzpicture}%
    } %
    \subfloat[SVR w/ Polynomial Kernel (Degree 3)]{%
        \pgfplotstableread{fig/out_err/SVR_PolyKernel_learn.csv}\learncsv%
        \begin{tikzpicture}
    \begin{axis}[
        plot-learn-time
    ]
        \addplot+ [learn-time-color, mark=triangle*]
            table [
                x expr=\thisrow{train size}*100,
                y expr=\thisrow{fit time mean}*1000
            ]{\learncsv}; \label{fit-time}
        
        \addplot [name path=time upper, draw=none] 
            table [
                x expr=\thisrow{train size}*100,
                y expr=\thisrow{fit time mean}*1000+\thisrow{fit time std}*1000
            ] {\learncsv};
        \addplot [name path=time lower, draw=none] 
            table[
                x expr=\thisrow{train size}*100,
                y expr=\thisrow{fit time mean}*1000-\thisrow{fit time std}*1000
            ] {\learncsv};
        \addplot [fill=learn-time-color, opacity=0.25] fill between[of=time upper and time lower];
        
    \end{axis}
    
    \begin{axis}[
        plot-learn,
    ]
        \addplot+ [learn-train-color]
            table [
                x expr=\thisrow{train size}*100,
                y=train mean
            ]{\learncsv};
        \addplot+ [learn-test-color]
            table [
                x expr=\thisrow{train size}*100, 
                y=test mean
            ]{\learncsv};
        \legend{Train Score, Test Score}
        \addlegendimage{/pgfplots/refstyle=fit-time}\addlegendentry{Fit Time}
        
        \addplot [name path=train upper, draw=none] 
            table [
                x expr=\thisrow{train size}*100,
                y expr=\thisrow{train mean}+\thisrow{train std}
            ] {\learncsv};
        \addplot [name path=train lower, draw=none] 
            table[
                x expr=\thisrow{train size}*100,
                y expr=\thisrow{train mean}-\thisrow{train std}
            ] {\learncsv};
        \addplot [fill=learn-train-color, opacity=0.25] fill between[of=train upper and train lower];
        
        \addplot [name path=train upper, draw=none] 
            table [
                x expr=\thisrow{train size}*100,
                y expr=\thisrow{test mean}+\thisrow{test std}
            ] {\learncsv};
        \addplot [name path=train lower, draw=none] 
            table[
                x expr=\thisrow{train size}*100,
                y expr=\thisrow{test mean}-\thisrow{test std}
            ] {\learncsv};
        \addplot [fill=learn-test-color, opacity=0.25] fill between[of=train upper and train lower];
    \end{axis}
\end{tikzpicture}%
    }%
    \subfloat[SVR w/ RBF Kernel]{%
        \pgfplotstableread{fig/out_err/SVR_RBFKernel_learn.csv}\learncsv%
        \begin{tikzpicture}
    \begin{axis}[
        plot-learn-time
    ]
        \addplot+ [learn-time-color, mark=triangle*]
            table [
                x expr=\thisrow{train size}*100,
                y expr=\thisrow{fit time mean}*1000
            ]{\learncsv}; \label{fit-time}
        
        \addplot [name path=time upper, draw=none] 
            table [
                x expr=\thisrow{train size}*100,
                y expr=\thisrow{fit time mean}*1000+\thisrow{fit time std}*1000
            ] {\learncsv};
        \addplot [name path=time lower, draw=none] 
            table[
                x expr=\thisrow{train size}*100,
                y expr=\thisrow{fit time mean}*1000-\thisrow{fit time std}*1000
            ] {\learncsv};
        \addplot [fill=learn-time-color, opacity=0.25] fill between[of=time upper and time lower];
        
    \end{axis}
    
    \begin{axis}[
        plot-learn,
    ]
        \addplot+ [learn-train-color]
            table [
                x expr=\thisrow{train size}*100,
                y=train mean
            ]{\learncsv};
        \addplot+ [learn-test-color]
            table [
                x expr=\thisrow{train size}*100, 
                y=test mean
            ]{\learncsv};
        \legend{Train Score, Test Score}
        \addlegendimage{/pgfplots/refstyle=fit-time}\addlegendentry{Fit Time}
        
        \addplot [name path=train upper, draw=none] 
            table [
                x expr=\thisrow{train size}*100,
                y expr=\thisrow{train mean}+\thisrow{train std}
            ] {\learncsv};
        \addplot [name path=train lower, draw=none] 
            table[
                x expr=\thisrow{train size}*100,
                y expr=\thisrow{train mean}-\thisrow{train std}
            ] {\learncsv};
        \addplot [fill=learn-train-color, opacity=0.25] fill between[of=train upper and train lower];
        
        \addplot [name path=train upper, draw=none] 
            table [
                x expr=\thisrow{train size}*100,
                y expr=\thisrow{test mean}+\thisrow{test std}
            ] {\learncsv};
        \addplot [name path=train lower, draw=none] 
            table[
                x expr=\thisrow{train size}*100,
                y expr=\thisrow{test mean}-\thisrow{test std}
            ] {\learncsv};
        \addplot [fill=learn-test-color, opacity=0.25] fill between[of=train upper and train lower];
    \end{axis}
\end{tikzpicture}%
    }%
    \caption{Learning curve and fit time predicting the Output Failure (cross validation = 10).}
    \label{fig:out-err-learning-curve}
\end{figure*}

\subsection{Evaluated Regression Models}

Several machine learning models were used to predict the two described failure classes, Output and Application Failure. Therefore, the data obtained from the fault injection campaign and the extracted flip-flop features form the training and test data set. All evaluated models are implemented using Python's scikit-learn Machine Learning framework~\cite{pedregosa_scikit-learn_2011}. 

Before learning the models the feature set is standardised, by removing the mean and scaling to unit variance. Further, since the FDR factors are within the range of 0 to 1, the predicted values are clipped to expected output range. For the hyperparameter search and evaluation a cross validation fold of 10 and a training size of 50\,\% are used.

\begin{figure*}[htbp]
    \centering
    \subfloat[Linear Least Squares]{%
        \pgfplotstableread{fig/app_err/LinearRegression_example.csv}\examplecsv%
        \begin{tikzpicture}
    \begin{axis}[
        plot-example
    ]
        \addplot+ [example-true-color, opacity=0.75] 
            table [x=FF idx, y=test true]{\examplecsv};
        \addplot+ [color=example-predicted-color, opacity=0.75]
            table [x=FF idx, y=test predicted]{\examplecsv};
        \legend{True Test Values, Predicted Test Values}
    \end{axis}
\end{tikzpicture}%
    }%
    \subfloat[$k$-Nearest Neighbors]{%
        \pgfplotstableread{fig/app_err/KNeighborsRegressor_example.csv}\examplecsv%
        \begin{tikzpicture}
    \begin{axis}[
        plot-example
    ]
        \addplot+ [example-true-color, opacity=0.75] 
            table [x=FF idx, y=test true]{\examplecsv};
        \addplot+ [color=example-predicted-color, opacity=0.75]
            table [x=FF idx, y=test predicted]{\examplecsv};
        \legend{True Test Values, Predicted Test Values}
    \end{axis}
\end{tikzpicture}%
    }%
    \subfloat[Decision Tree]{%
        \pgfplotstableread{fig/app_err/DecisionTreeRegressor_example.csv}\examplecsv%
        \begin{tikzpicture}
    \begin{axis}[
        plot-example
    ]
        \addplot+ [example-true-color, opacity=0.75] 
            table [x=FF idx, y=test true]{\examplecsv};
        \addplot+ [color=example-predicted-color, opacity=0.75]
            table [x=FF idx, y=test predicted]{\examplecsv};
        \legend{True Test Values, Predicted Test Values}
    \end{axis}
\end{tikzpicture}%
    }%
    
    \subfloat[Ridge w/ Polynomial Kernel (Degree 2)]{%
        \pgfplotstableread{fig/app_err/KernelRidgeRegressor_2DegreePolyKernel_example.csv}\examplecsv%
        \begin{tikzpicture}
    \begin{axis}[
        plot-example
    ]
        \addplot+ [example-true-color, opacity=0.75] 
            table [x=FF idx, y=test true]{\examplecsv};
        \addplot+ [color=example-predicted-color, opacity=0.75]
            table [x=FF idx, y=test predicted]{\examplecsv};
        \legend{True Test Values, Predicted Test Values}
    \end{axis}
\end{tikzpicture}%
    }%
    \subfloat[SVR w/ Polynomial Kernel (Degree 5)]{%
        \pgfplotstableread{fig/app_err/SVR_PolyKernel_example.csv}\examplecsv%
        \begin{tikzpicture}
    \begin{axis}[
        plot-example
    ]
        \addplot+ [example-true-color, opacity=0.75] 
            table [x=FF idx, y=test true]{\examplecsv};
        \addplot+ [color=example-predicted-color, opacity=0.75]
            table [x=FF idx, y=test predicted]{\examplecsv};
        \legend{True Test Values, Predicted Test Values}
    \end{axis}
\end{tikzpicture}%
    }%
    \subfloat[SVR w/ RBF Kernel]{%
        \pgfplotstableread{fig/app_err/SVR_RBFKernel_example.csv}\examplecsv%
        \begin{tikzpicture}
    \begin{axis}[
        plot-example
    ]
        \addplot+ [example-true-color, opacity=0.75] 
            table [x=FF idx, y=test true]{\examplecsv};
        \addplot+ [color=example-predicted-color, opacity=0.75]
            table [x=FF idx, y=test predicted]{\examplecsv};
        \legend{True Test Values, Predicted Test Values}
    \end{axis}
\end{tikzpicture}%
    }%
    \caption{Prediction of the Application Failure for one test data fold (training size = 50\%).}
    \label{fig:app-err-example}
\end{figure*}
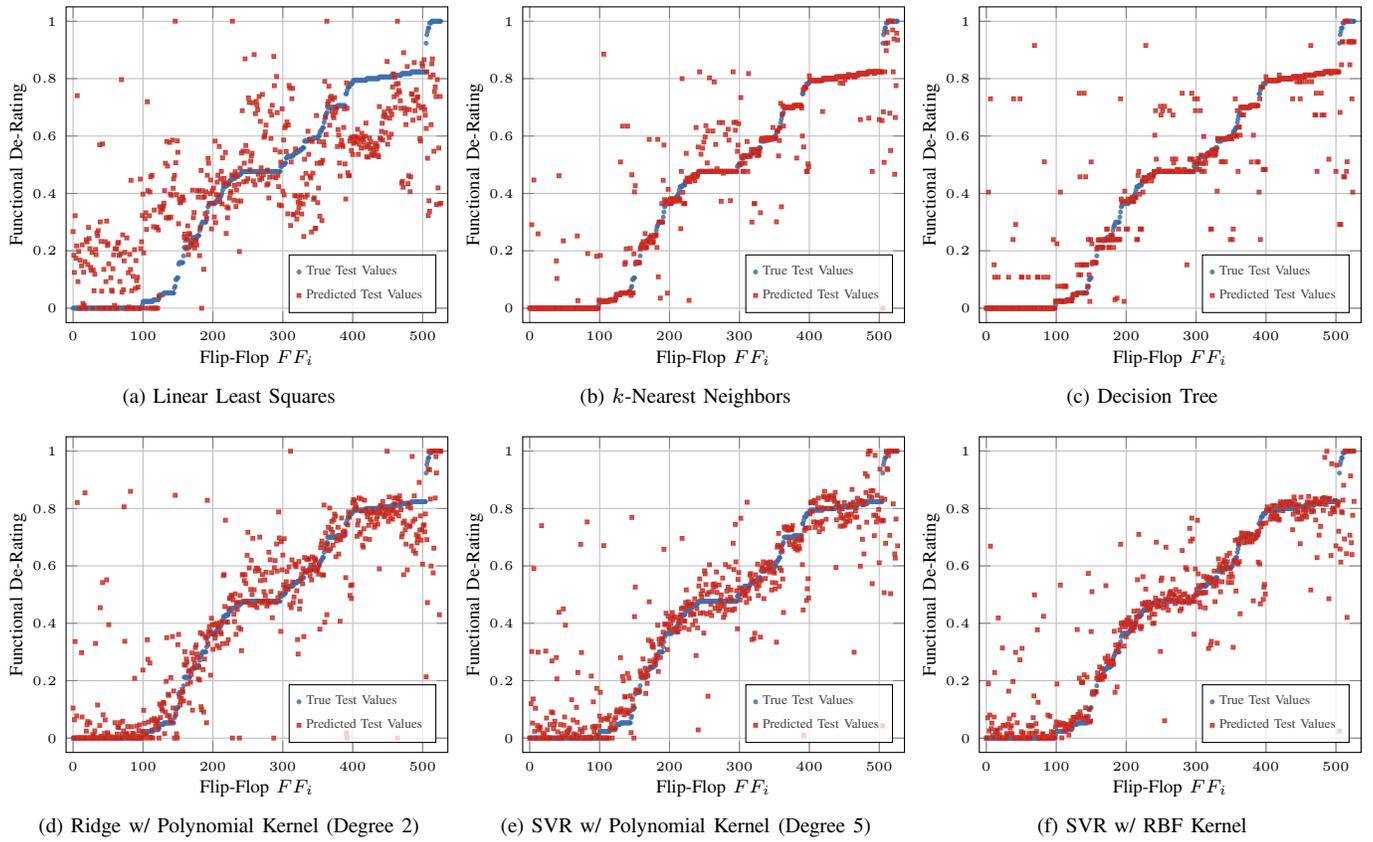

The investigated models are briefly described in the following (for a more detailed description see \cite{murphy_machine_2012} for example) and 
the prediction performances for the two failure classes are given in Table~\ref{tab:estimation_restults}. For a selection of models
the prediction of one test data fold is shown in Fig.\ref{fig:out-err-example} for the Output Failure and Fig.\ref{fig:app-err-example} for the Application Failure. Further, Fig.\ref{fig:out-err-learning-curve} and Fig.\ref{fig:app-err-learning-curve} show the learning curves, which describes the performance of the model for different training sizes.

\begin{table*}[htbp]
    \centering
    \caption{%
        Performance Results for the Evaluated Regression Models %
        (Cross Validation = 10, Training Size = 50\,\%)%
    }
    \label{tab:estimation_restults}
    \vspace{-15pt}
    \subfloat[Output Failure]{
        \scriptsize{
        \begin{tabular}{%
            l *{5}{S[table-format=1.3]} @{\hspace{15pt}}%
            S[table-format=6.0]%
            S[table-format=4.0]%
            *{2}{S[table-format=1.3]}%
        }
        \toprule
            \heading{Regression Model} &
            \heading{MAE} & \heading{MAX} & \heading{RMSE} & 
            \heading{EV} & \heading{$R^2$} &
            \heading{\shortstack{Hyperparameter \\ Combinations}} &
            \heading{\shortstack{Training \\ Time / s}} & 
            \heading{\shortstack{Fit \\ Time / s}} & 
            \heading{\shortstack{Prediction \\ Time / s}}\\
        \midrule
            Linear Least Squares &
				0.12 & 0.817 & 0.17 & 0.671 & 0.67 &
				\heading{-} & \heading{-} & 0.004 & 0.002\\
            $k$-Nearest Neighbors &
				0.045 & 0.714 & 0.108 & 0.867 & 0.866 &
				4500 & 1730 & 0.002 & 0.013\\
            Decision Tree &
				0.048 & 0.802 & 0.123 & 0.824 & 0.823 &
				250000 & 1765 & 0.004 & 0.001\\
            Ridge w/ Linear Kernel &
				0.119 & 0.759 & 0.167 & 0.681 & 0.68 &
				50000 & 1221 & 0.007 & 0.002\\
            Ridge w/ Polynomial Kernel &
				0.077 & 0.789 & 0.132 & 0.801 & 0.8 &
				50000 & 1375 & 0.008 & 0.002\\
            Ridge w/ RBF Kernel &
				0.074 & 0.748 & 0.133 & 0.798 & 0.797 &
				50000 & 1836 & 0.009 & 0.004\\
            Ridge w/ Sigmoid Kernel &
				0.143 & 0.822 & 0.212 & 0.489 & 0.486 &
				20000 & 1689 & 0.038 & 0.003\\
            SVR w/ linear kernel &
				0.114 & 0.803 & 0.165 & 0.69 & 0.689 &
				50000 & 913 & 0.041 & 0.006\\
            SVR w/ polynomial kernel &
				0.063 & 0.73 & 0.112 & 0.857 & 0.856 &
				15000 & 1447 & 0.018 & 0.005\\
            SVR w/ RBF Kernel &
				0.053 & 0.648 & 0.102 & 0.88 & 0.879 &
				50000 & 1877 & 0.023 & 0.008\\
            SVR w/ Sigmoid kernel &
				0.122 & 0.747 & 0.172 & 0.663 & 0.661 &
				75000 & 2131 & 0.019 & 0.013\\
        \bottomrule
        \end{tabular}
        }
    }
    
    \subfloat[Application Failure]{
        \scriptsize{
        \begin{tabular}{%
            l *{5}{S[table-format=1.3]} @{\hspace{15pt}}%
            S[table-format=6.0]%
            S[table-format=4.0]%
            *{2}{S[table-format=1.3]}%
        }
        \toprule
            \heading{Regression Model} &
            \heading{MAE} & \heading{MAX} & \heading{RMSE} & 
            \heading{EV} & \heading{$R^2$} & 
            \heading{\shortstack{Hyperparameter \\ Combinations}} &
            \heading{\shortstack{Training \\ Time / s}} & 
            \heading{\shortstack{Fit \\ Time / s}} & 
            \heading{\shortstack{Prediction \\ Time / s}}\\
        \midrule
            Linear Least Squares &
                0.161 & 0.872 & 0.213 & 0.54 & 0.539 &
                \heading{-} & \heading{-} & 0.003 & 0.002 \\
            $k$-Nearest Neighbors &
                0.047 & 0.895 & 0.124 & 0.844 & 0.843 &
                4500 & 1721 & 0.002 & 0.015 \\
            Decision Tree &
                0.056 & 0.919 & 0.137 & 0.809 & 0.808 &
                250000 & 1752 & 0.003 & 0.001 \\
            Ridge w/ Linear Kernel &
                0.161 & 0.872 & 0.213 & 0.54 & 0.539 &
                50000 & 1334 & 0.008 & 0.002 \\
            Ridge w/ Polynomial Kernel &
                0.08 & 0.896 & 0.142 & 0.793 & 0.793 &
                50000 & 1464 & 0.007 & 0.002 \\
            Ridge w/ RBF Kernel &
                0.079 & 0.874 & 0.142 & 0.795 & 0.794 &
                50000 & 1952 & 0.008 & 0.004 \\
            Ridge w/ Sigmoid Kernel &
                0.187 & 0.991 & 0.254 & 0.339 & 0.332 &
                20000 & 2291 & 0.04 & 0.003 \\
            SVR w/ Linear Kernel &
                0.167 & 0.858 & 0.218 & 0.519 & 0.518 &
                50000 & 2065 & 0.022 & 0.007 \\
            SVR w/ Polynomial Kernel &
                0.071 & 0.904 & 0.135 & 0.816 & 0.815 &
                15000 & 1944 & 0.026 & 0.006 \\
            SVR w/ RBF Kernel &
                0.06 & 0.86 & 0.123 & 0.846 & 0.846 &
                50000 & 2067 & 0.024 & 0.008 \\
            SVR w/ Sigmoid Kernel &
                0.234 & 0.717 & 0.275 & 0.24 & 0.23 &
                300000 & 2024 & 0.003 & 0.002 \\
        \bottomrule
        \end{tabular}
        }
    }
\end{table*}

\subsubsection{Linear Least Squares Regression}

The Linear Least Squares algorithm, is based on a linear model. The target output variable is represented as a linear combination of the input feature variables. Thereby, the algorithm targets to minimise the sum of squared residuals, the squared sum of the difference between the true value in the training dataset and the predicted value by the linear approximation.

\subsubsection{k-Nearest Neighbors Regression}

The $k$-Nearest Neighbor ($k$-NN) algorithm exploits feature similarity to predict values of new data points. During the training phase the training data set is only indexed and stored into a database. Then, the value of a new data point is predicted based on how closely it resembles to the points in the training set. A weighted average of the $k$-nearest neighbors is used to predict the value, where the weight is calculated by the inverse of the distances and the distance itself can be any metric measure, such as the Manhattan or Euclidean distance. Hence, the model hyperparameters are $k$ and the distance metrics.

During the hyperparameter optimisation it was found that the model performed the best with $k=3$ and Manhattan distance as metric measure for both failure classes.

\subsubsection{Decision Tree Regression}

Decision Trees in Machine Learning are models which recursively partitioning the input feature space by inferring simple decision rules from the training data. This is usually represented by a tree structure where the decision rules are defined in the branches of the tree and the leaves contain the trained value. In general, the deeper the tree, the more different decision rules it has which results in a more complex model. However, this can also lead to over-complex trees that do not generalise the training data, also called overfitting.

The considered hyperparameters for this model are controlling the structure of the tree, such as the maximum depth, the maximum number of leaf nodes and the balance of the tree. Further parameters defined by the framework are set to their default values.

The hyperparameter optimisation has shown that the model performs at best for both failure classes when the tree structure is not restricted (no maximum for the depth and number of leaf nodes is set and no restriction to balance the tree). 

\subsubsection{Kernel Ridge Regression}

The Ridge regression algorithm is modifying the ordinary least squares regression by imposing a penalty on the size of the coefficients (regularisation). Another advantage of the Ridge regression is, that it can be extended to use kernel functions. These functions perform a transformation of the input values and map them to a higher dimensional space. Thus, it is possible to learn a linear function in the space induced by the respective kernel. For non-linear kernels, this corresponds to a non-linear function in the original space. This is useful for regression problems which cannot adequately be described by linear models.

The hyperparameter of the model are the regularisation hyperparameter $\alpha$, $\gamma$ to control the kernel function, $d$ to define the degree of the polynomial kernel and the independent term $r$ in the polynomial and sigmoid kernel.

The best performance when using a Ridge regression with a polynomial kernel was found with degree of $d=2$ for both failure classes.

\subsubsection{Support Vector Regression}

The Support Vector Regression (SVR) is similar to the Ridge Regression where the loss function is modified in such a way that predictions only depend on a subset of the training data, known as support vectors. The goal is to find a function where each training data points within an $\varepsilon$-tube are not penalised, and at the same time is as flat as possible. SVR can also operate with the kernel trick and the framework provides the same kernels as for the Ridge Regression. 
The model defines several hyperparameters, such as the penalty factor~$C$, the size of the $\varepsilon$-tube, $\gamma$ to control the kernel function, $d$ to define the degree of the polynomial kernel and the independent term $r$ in the polynomial and sigmoid kernel.

The SVR with a polynomial kernel operated the best when a degree of $d=3$ for Output Failure and $d=5$ for the Application Failure was chosen.

\subsection{Comparison and Discussion}

\subsubsection{Prediction Performance}

Comparing the general performance of the different models shown in Tab.~\ref{tab:estimation_restults}, it can be seen that the Linear Least Squares regression and the regression models using linear kernels are rated the worst. The Decision Tree regression, instance-based $k$-NN algorithm, as well as the kernel-based algorithms using a non-linear kernel performing much better (except by using the Sigmoid kernel). This suggests that the features, used for the prediction, are not linear dependent to the target variable, the Functional De-Rating factor.

The prediction performance of the Output Failure compared to the Application Failure is better in all cases, especially for the linear models. This could be explained by the additional complexity the models have to learn to predict the failure on the application level, but also indicates missing features which are more representative for the application. However, looking at the non-linear models the difference is not very significant, which demonstrates that these models are also able to learn the higher complexity. Further, increasing the number of features can also decrease the performance due to the curse of dimensionality\cite{trunk_problem_1979}.

The learning curves in Fig.~\ref{fig:out-err-learning-curve} and Fig.\ref{fig:app-err-learning-curve}, show that the performance does not significantly improve with training sizes higher than 50\,\%. This means, by using the proposed method in a fault injection campaign, the time needed to obtain a detailed list of Functional De-Rating factors can be reduced by half. The cost can even be reduced further, in exchange of a slight reduction in accuracy of about $10\,\%$. Thus, a more aggressive optimisation, a cost reduction up-to $5\times$, can be achieved.

\subsubsection{Training and Prediction Time}

The fit and prediction time of a model defines how many hyperparameter combinations can be tested in order to find the best performance in a reasonable amount of time. In this paper, the number of hyperparamter combinations was chosen such that the total needed training time is about $\SI{30}{minutes}$, in order to have a substantial advantage to the fault injection simulation. Thus, the total time to extract the features and train a model would be about 3\,\% of the total time needed to perform the full fault injection campaign. In addition, it should be noted that the used machine learning framework does not require any licenses and offers several functions to parallelize the computation on clusters. Hence, testing the hyperparameter combinations can be accelerated more easily than the accelerating the fault injection campaign.

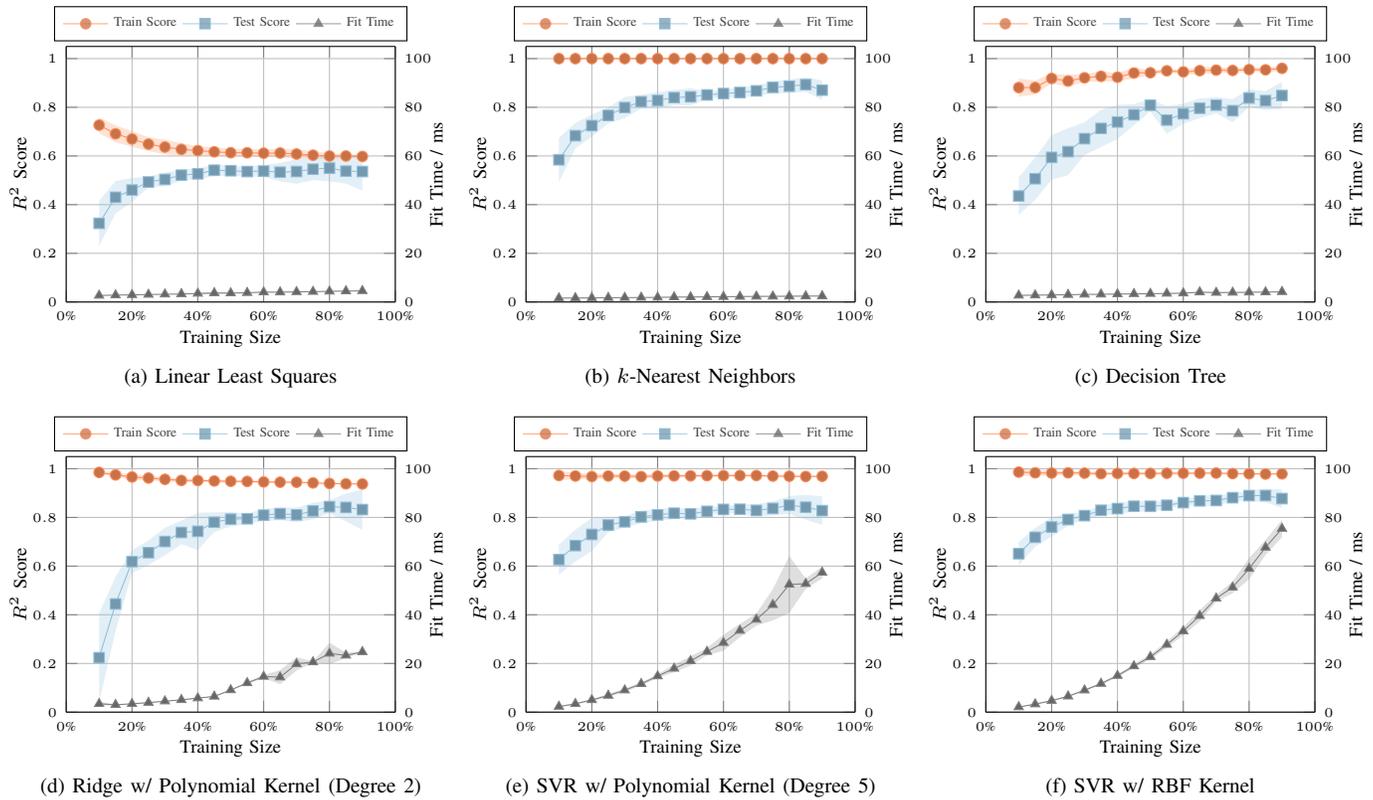
\begin{figure*}[htbp]
    \vspace{-20pt}%
    \centering%
    \subfloat[Linear Least Squares]{%
        \pgfplotstableread{fig/app_err/LinearRegression_learn.csv}\learncsv%
        \begin{tikzpicture}
    \begin{axis}[
        plot-learn-time
    ]
        \addplot+ [learn-time-color, mark=triangle*]
            table [
                x expr=\thisrow{train size}*100,
                y expr=\thisrow{fit time mean}*1000
            ]{\learncsv}; \label{fit-time}
        
        \addplot [name path=time upper, draw=none] 
            table [
                x expr=\thisrow{train size}*100,
                y expr=\thisrow{fit time mean}*1000+\thisrow{fit time std}*1000
            ] {\learncsv};
        \addplot [name path=time lower, draw=none] 
            table[
                x expr=\thisrow{train size}*100,
                y expr=\thisrow{fit time mean}*1000-\thisrow{fit time std}*1000
            ] {\learncsv};
        \addplot [fill=learn-time-color, opacity=0.25] fill between[of=time upper and time lower];
        
    \end{axis}
    
    \begin{axis}[
        plot-learn,
    ]
        \addplot+ [learn-train-color]
            table [
                x expr=\thisrow{train size}*100,
                y=train mean
            ]{\learncsv};
        \addplot+ [learn-test-color]
            table [
                x expr=\thisrow{train size}*100, 
                y=test mean
            ]{\learncsv};
        \legend{Train Score, Test Score}
        \addlegendimage{/pgfplots/refstyle=fit-time}\addlegendentry{Fit Time}
        
        \addplot [name path=train upper, draw=none] 
            table [
                x expr=\thisrow{train size}*100,
                y expr=\thisrow{train mean}+\thisrow{train std}
            ] {\learncsv};
        \addplot [name path=train lower, draw=none] 
            table[
                x expr=\thisrow{train size}*100,
                y expr=\thisrow{train mean}-\thisrow{train std}
            ] {\learncsv};
        \addplot [fill=learn-train-color, opacity=0.25] fill between[of=train upper and train lower];
        
        \addplot [name path=train upper, draw=none] 
            table [
                x expr=\thisrow{train size}*100,
                y expr=\thisrow{test mean}+\thisrow{test std}
            ] {\learncsv};
        \addplot [name path=train lower, draw=none] 
            table[
                x expr=\thisrow{train size}*100,
                y expr=\thisrow{test mean}-\thisrow{test std}
            ] {\learncsv};
        \addplot [fill=learn-test-color, opacity=0.25] fill between[of=train upper and train lower];
    \end{axis}
\end{tikzpicture}%
    }%
    \subfloat[$k$-Nearest Neighbors]{%
        \pgfplotstableread{fig/app_err/KNeighborsRegressor_learn.csv}\learncsv%
        \begin{tikzpicture}
    \begin{axis}[
        plot-learn-time
    ]
        \addplot+ [learn-time-color, mark=triangle*]
            table [
                x expr=\thisrow{train size}*100,
                y expr=\thisrow{fit time mean}*1000
            ]{\learncsv}; \label{fit-time}
        
        \addplot [name path=time upper, draw=none] 
            table [
                x expr=\thisrow{train size}*100,
                y expr=\thisrow{fit time mean}*1000+\thisrow{fit time std}*1000
            ] {\learncsv};
        \addplot [name path=time lower, draw=none] 
            table[
                x expr=\thisrow{train size}*100,
                y expr=\thisrow{fit time mean}*1000-\thisrow{fit time std}*1000
            ] {\learncsv};
        \addplot [fill=learn-time-color, opacity=0.25] fill between[of=time upper and time lower];
        
    \end{axis}
    
    \begin{axis}[
        plot-learn,
    ]
        \addplot+ [learn-train-color]
            table [
                x expr=\thisrow{train size}*100,
                y=train mean
            ]{\learncsv};
        \addplot+ [learn-test-color]
            table [
                x expr=\thisrow{train size}*100, 
                y=test mean
            ]{\learncsv};
        \legend{Train Score, Test Score}
        \addlegendimage{/pgfplots/refstyle=fit-time}\addlegendentry{Fit Time}
        
        \addplot [name path=train upper, draw=none] 
            table [
                x expr=\thisrow{train size}*100,
                y expr=\thisrow{train mean}+\thisrow{train std}
            ] {\learncsv};
        \addplot [name path=train lower, draw=none] 
            table[
                x expr=\thisrow{train size}*100,
                y expr=\thisrow{train mean}-\thisrow{train std}
            ] {\learncsv};
        \addplot [fill=learn-train-color, opacity=0.25] fill between[of=train upper and train lower];
        
        \addplot [name path=train upper, draw=none] 
            table [
                x expr=\thisrow{train size}*100,
                y expr=\thisrow{test mean}+\thisrow{test std}
            ] {\learncsv};
        \addplot [name path=train lower, draw=none] 
            table[
                x expr=\thisrow{train size}*100,
                y expr=\thisrow{test mean}-\thisrow{test std}
            ] {\learncsv};
        \addplot [fill=learn-test-color, opacity=0.25] fill between[of=train upper and train lower];
    \end{axis}
\end{tikzpicture}%
    }%
    \subfloat[Decision Tree]{%
        \pgfplotstableread{fig/app_err/DecisionTreeRegressor_learn.csv}\learncsv%
        \begin{tikzpicture}
    \begin{axis}[
        plot-learn-time
    ]
        \addplot+ [learn-time-color, mark=triangle*]
            table [
                x expr=\thisrow{train size}*100,
                y expr=\thisrow{fit time mean}*1000
            ]{\learncsv}; \label{fit-time}
        
        \addplot [name path=time upper, draw=none] 
            table [
                x expr=\thisrow{train size}*100,
                y expr=\thisrow{fit time mean}*1000+\thisrow{fit time std}*1000
            ] {\learncsv};
        \addplot [name path=time lower, draw=none] 
            table[
                x expr=\thisrow{train size}*100,
                y expr=\thisrow{fit time mean}*1000-\thisrow{fit time std}*1000
            ] {\learncsv};
        \addplot [fill=learn-time-color, opacity=0.25] fill between[of=time upper and time lower];
        
    \end{axis}
    
    \begin{axis}[
        plot-learn,
    ]
        \addplot+ [learn-train-color]
            table [
                x expr=\thisrow{train size}*100,
                y=train mean
            ]{\learncsv};
        \addplot+ [learn-test-color]
            table [
                x expr=\thisrow{train size}*100, 
                y=test mean
            ]{\learncsv};
        \legend{Train Score, Test Score}
        \addlegendimage{/pgfplots/refstyle=fit-time}\addlegendentry{Fit Time}
        
        \addplot [name path=train upper, draw=none] 
            table [
                x expr=\thisrow{train size}*100,
                y expr=\thisrow{train mean}+\thisrow{train std}
            ] {\learncsv};
        \addplot [name path=train lower, draw=none] 
            table[
                x expr=\thisrow{train size}*100,
                y expr=\thisrow{train mean}-\thisrow{train std}
            ] {\learncsv};
        \addplot [fill=learn-train-color, opacity=0.25] fill between[of=train upper and train lower];
        
        \addplot [name path=train upper, draw=none] 
            table [
                x expr=\thisrow{train size}*100,
                y expr=\thisrow{test mean}+\thisrow{test std}
            ] {\learncsv};
        \addplot [name path=train lower, draw=none] 
            table[
                x expr=\thisrow{train size}*100,
                y expr=\thisrow{test mean}-\thisrow{test std}
            ] {\learncsv};
        \addplot [fill=learn-test-color, opacity=0.25] fill between[of=train upper and train lower];
    \end{axis}
\end{tikzpicture}%
    }%
    
    \subfloat[Ridge w/ Polynomial Kernel (Degree 2)]{%
        \pgfplotstableread{fig/app_err/KernelRidgeRegressor_2DegreePolyKernel_learn.csv}\learncsv%
        \begin{tikzpicture}
    \begin{axis}[
        plot-learn-time
    ]
        \addplot+ [learn-time-color, mark=triangle*]
            table [
                x expr=\thisrow{train size}*100,
                y expr=\thisrow{fit time mean}*1000
            ]{\learncsv}; \label{fit-time}
        
        \addplot [name path=time upper, draw=none] 
            table [
                x expr=\thisrow{train size}*100,
                y expr=\thisrow{fit time mean}*1000+\thisrow{fit time std}*1000
            ] {\learncsv};
        \addplot [name path=time lower, draw=none] 
            table[
                x expr=\thisrow{train size}*100,
                y expr=\thisrow{fit time mean}*1000-\thisrow{fit time std}*1000
            ] {\learncsv};
        \addplot [fill=learn-time-color, opacity=0.25] fill between[of=time upper and time lower];
        
    \end{axis}
    
    \begin{axis}[
        plot-learn,
    ]
        \addplot+ [learn-train-color]
            table [
                x expr=\thisrow{train size}*100,
                y=train mean
            ]{\learncsv};
        \addplot+ [learn-test-color]
            table [
                x expr=\thisrow{train size}*100, 
                y=test mean
            ]{\learncsv};
        \legend{Train Score, Test Score}
        \addlegendimage{/pgfplots/refstyle=fit-time}\addlegendentry{Fit Time}
        
        \addplot [name path=train upper, draw=none] 
            table [
                x expr=\thisrow{train size}*100,
                y expr=\thisrow{train mean}+\thisrow{train std}
            ] {\learncsv};
        \addplot [name path=train lower, draw=none] 
            table[
                x expr=\thisrow{train size}*100,
                y expr=\thisrow{train mean}-\thisrow{train std}
            ] {\learncsv};
        \addplot [fill=learn-train-color, opacity=0.25] fill between[of=train upper and train lower];
        
        \addplot [name path=train upper, draw=none] 
            table [
                x expr=\thisrow{train size}*100,
                y expr=\thisrow{test mean}+\thisrow{test std}
            ] {\learncsv};
        \addplot [name path=train lower, draw=none] 
            table[
                x expr=\thisrow{train size}*100,
                y expr=\thisrow{test mean}-\thisrow{test std}
            ] {\learncsv};
        \addplot [fill=learn-test-color, opacity=0.25] fill between[of=train upper and train lower];
    \end{axis}
\end{tikzpicture}%
    } %
    \subfloat[SVR w/ Polynomial Kernel (Degree 5)]{%
        \pgfplotstableread{fig/app_err/SVR_PolyKernel_learn.csv}\learncsv%
        \begin{tikzpicture}
    \begin{axis}[
        plot-learn-time
    ]
        \addplot+ [learn-time-color, mark=triangle*]
            table [
                x expr=\thisrow{train size}*100,
                y expr=\thisrow{fit time mean}*1000
            ]{\learncsv}; \label{fit-time}
        
        \addplot [name path=time upper, draw=none] 
            table [
                x expr=\thisrow{train size}*100,
                y expr=\thisrow{fit time mean}*1000+\thisrow{fit time std}*1000
            ] {\learncsv};
        \addplot [name path=time lower, draw=none] 
            table[
                x expr=\thisrow{train size}*100,
                y expr=\thisrow{fit time mean}*1000-\thisrow{fit time std}*1000
            ] {\learncsv};
        \addplot [fill=learn-time-color, opacity=0.25] fill between[of=time upper and time lower];
        
    \end{axis}
    
    \begin{axis}[
        plot-learn,
    ]
        \addplot+ [learn-train-color]
            table [
                x expr=\thisrow{train size}*100,
                y=train mean
            ]{\learncsv};
        \addplot+ [learn-test-color]
            table [
                x expr=\thisrow{train size}*100, 
                y=test mean
            ]{\learncsv};
        \legend{Train Score, Test Score}
        \addlegendimage{/pgfplots/refstyle=fit-time}\addlegendentry{Fit Time}
        
        \addplot [name path=train upper, draw=none] 
            table [
                x expr=\thisrow{train size}*100,
                y expr=\thisrow{train mean}+\thisrow{train std}
            ] {\learncsv};
        \addplot [name path=train lower, draw=none] 
            table[
                x expr=\thisrow{train size}*100,
                y expr=\thisrow{train mean}-\thisrow{train std}
            ] {\learncsv};
        \addplot [fill=learn-train-color, opacity=0.25] fill between[of=train upper and train lower];
        
        \addplot [name path=train upper, draw=none] 
            table [
                x expr=\thisrow{train size}*100,
                y expr=\thisrow{test mean}+\thisrow{test std}
            ] {\learncsv};
        \addplot [name path=train lower, draw=none] 
            table[
                x expr=\thisrow{train size}*100,
                y expr=\thisrow{test mean}-\thisrow{test std}
            ] {\learncsv};
        \addplot [fill=learn-test-color, opacity=0.25] fill between[of=train upper and train lower];
    \end{axis}
\end{tikzpicture}%
    }%
    \subfloat[SVR w/ RBF Kernel]{%
        \pgfplotstableread{fig/app_err/SVR_RBFKernel_learn.csv}\learncsv%
        \begin{tikzpicture}
    \begin{axis}[
        plot-learn-time
    ]
        \addplot+ [learn-time-color, mark=triangle*]
            table [
                x expr=\thisrow{train size}*100,
                y expr=\thisrow{fit time mean}*1000
            ]{\learncsv}; \label{fit-time}
        
        \addplot [name path=time upper, draw=none] 
            table [
                x expr=\thisrow{train size}*100,
                y expr=\thisrow{fit time mean}*1000+\thisrow{fit time std}*1000
            ] {\learncsv};
        \addplot [name path=time lower, draw=none] 
            table[
                x expr=\thisrow{train size}*100,
                y expr=\thisrow{fit time mean}*1000-\thisrow{fit time std}*1000
            ] {\learncsv};
        \addplot [fill=learn-time-color, opacity=0.25] fill between[of=time upper and time lower];
        
    \end{axis}
    
    \begin{axis}[
        plot-learn,
    ]
        \addplot+ [learn-train-color]
            table [
                x expr=\thisrow{train size}*100,
                y=train mean
            ]{\learncsv};
        \addplot+ [learn-test-color]
            table [
                x expr=\thisrow{train size}*100, 
                y=test mean
            ]{\learncsv};
        \legend{Train Score, Test Score}
        \addlegendimage{/pgfplots/refstyle=fit-time}\addlegendentry{Fit Time}
        
        \addplot [name path=train upper, draw=none] 
            table [
                x expr=\thisrow{train size}*100,
                y expr=\thisrow{train mean}+\thisrow{train std}
            ] {\learncsv};
        \addplot [name path=train lower, draw=none] 
            table[
                x expr=\thisrow{train size}*100,
                y expr=\thisrow{train mean}-\thisrow{train std}
            ] {\learncsv};
        \addplot [fill=learn-train-color, opacity=0.25] fill between[of=train upper and train lower];
        
        \addplot [name path=train upper, draw=none] 
            table [
                x expr=\thisrow{train size}*100,
                y expr=\thisrow{test mean}+\thisrow{test std}
            ] {\learncsv};
        \addplot [name path=train lower, draw=none] 
            table[
                x expr=\thisrow{train size}*100,
                y expr=\thisrow{test mean}-\thisrow{test std}
            ] {\learncsv};
        \addplot [fill=learn-test-color, opacity=0.25] fill between[of=train upper and train lower];
    \end{axis}
\end{tikzpicture}%
    }%
    \caption{Learning curve and fit time predicting the Application Failure (cross validation = 10).}
    \label{fig:app-err-learning-curve}
\end{figure*}

Looking at the fit time in relation to the training size, shown in Fig.~\ref{fig:out-err-learning-curve} and Fig.\ref{fig:app-err-learning-curve}, it can be observed that with more training data the time needed to fit a model is increasing. Especially, the Ridge algorithm and SVR are known for a quadratic dependency of the training data. In contrary, the fit time of the $k$-NN and Decision Tree algorithm is increasing linearly with the training data~\cite{murphy_machine_2012}. This might make them more suitable to learn very large circuits with a high number of flip-flops.

\section{Conclusion and Future Work}
\label{sec:conclusion}

In this paper, machine learning techniques were used to assist the Functional Failure analysis of complex circuits. The presented methodology reduces the computational cost to determine the Functional De-Rating factors of the circuit's sequential logic. The aim is to predict factors per individual instances, which is particularly difficult to obtain using classical approaches such as clustering, selective fault simulation or fault universe compaction techniques. 

The methodology was applied in a practical example where several machine learning models were evaluated, predicting two different failure classes, first, predicting the propagation of a fault to the primary output and second, predicting the rate of a fault leading to a functional failure. The performance comparison has shown that linear models are not suitable to fit the problem. Much better performances were achieved with the instance-based $k$-NN model, the Decision Tree regression or kernel-based algorithm with non-linear kernels and training sizes of 20\% to 50\%. This means, the cost can be reduced of a fault injection campaign can be reduced by a factor of 2 up to 5 times in compared to a classical statistical fault injection campaign.

The comparison between the two failure classes has shown that the machine learning models are better in predicting the fault propagation to the output better than the functional failure rate, although the difference is not significant for non-linear models. This indicates that new features should be considered which characterise the functional behaviour more accurately. Further, the  value of each feature should be evaluated separately, in order to reduce the dimension which might have a positive effect on the performance, as well as prediction and training time.

\bibliographystyle{IEEEtran}
\bibliography{IEEEabrv,bib/SELSE2019.bib}

\end{document}